\documentclass[
    reprint, 
    amsmath, 
    amssymb, 
    aps, 
    prd, 
    superscriptaddress,
    nofootinbib
    ]{revtex4-2}

\usepackage{graphicx}

\usepackage{hyperref}

\usepackage{physics}

\usepackage{orcidlink}

\newcommand{\MM}{\mathcal{M}}
\newcommand{\tpeak}{t_\mathrm{peak}^{h_{22}}}

\newcommand{\CIT}{\affiliation{TAPIR, California Institute of Technology, Pasadena, CA 91125, USA}}
\newcommand{\CITLab}{\affiliation{LIGO Laboratory, California Institute of Technology, Pasadena, CA 91125, USA}}
\newcommand{\GAC}{\affiliation{Department of Physics, Gustavus Adolphus College, Saint Peter, MN 56082, USA}}

\begin{document}

\title{High-overtone ringdown fits: start time, no-hair tests, and correlations}

\author{Erin Coleman
\orcidlink{0009-0008-3927-4638}}
\email{erc@uchicago.edu}
\CITLab
\GAC

\author{Eliot Finch 
\orcidlink{0000-0002-1993-4263}}
\email{efinch@caltech.edu}
\CITLab
\CIT

\date{\today}

\begin{abstract}
Overtones are known to improve the performance of fits to the ringdown, both in numerical-relativity simulations and gravitational-wave observations.
Although the overtone frequencies are a concrete prediction of general relativity, it remains an open question whether they are excited to the extent that fits would suggest.
In this work, we take a pragmatic approach and investigate the practical utility of each additional overtone in extracting information from the ringdown. 
We look at the dependence of the ringdown start time on the number of overtones, and the feasibility of detecting deviations from general relativity in the ringdown frequencies.
We suggest that there is no clear ``maximum'' overtone, but rather the utility of each additional overtone decreases compared to the one before.
Finally, we perform Bayesian parameter estimation (as opposed to least-squares fits) to obtain posterior distributions on the overtone amplitudes and phases, allowing us to investigate their correlation structure.
Due to strong correlations it becomes increasingly hard to measure individual amplitudes and phases for the highest overtones.
However, we find that the joint measurement of overtone amplitudes (i.e., the correlation structure itself) is sensitive to the frequencies and decay times of even the highest overtones, possibly offering an avenue to perform consistency tests with general relativity.
\end{abstract}

\maketitle

\section{Introduction}\label{introduction}

A perturbed black hole (BH) radiates gravitational waves (GWs) as it settles to a final stationary state, in what is called the ringdown~\cite{Berti:2025hly}.
The ringdown waveform comprises a sum of quasinormal modes (QNMs); these are exponentially damped sinusoids whose discrete spectrum of complex frequencies $\omega_{\ell m n}(M_f, \chi_f)$ are uniquely determined by the final BH mass $M_f$ and dimensionless spin $\chi_f$.
The real part encodes the angular frequency and the negative imaginary part gives the inverse of the damping time:
\begin{equation}
    \omega_{\ell m n} = 2\pi f_{\ell m n} - i/\tau_{\ell m n}\,.
\end{equation}
The $\ell$ and $m$ are the polar and azimuthal indices associated with the spin-weighted spheroidal harmonics, which describe the angular dependence of the mode.
For each $(\ell,m)$ there exists an infinite set of modes labeled with an ``overtone'' index $n$.
These overtones are ordered according to their decay time, such that $n = 0$ is the longest lived ``fundamental'' mode, and overtones with $n > 0$ decay more quickly.\footnote{In general a fourth index is needed to distinguish between prograde and retrograde modes (or, alternatively, ``regular'' and ``mirror'' modes.) We will only be working with the prograde family of modes in this work and so we drop the fourth index.}

Despite being shorter lived than the fundamental modes, with the advent of numerical relativity (NR)~\cite{Pretorius:2005gq, Campanelli:2005dd, Baker:2005vv} it was soon found that including overtones in fits to the ringdown helped to accurately recover the remnant BH properties and allowed fits to be performed at earlier times (see \citet{Carullo:2025oms} for a succinct review). 
Early works~\cite{Dorband:2006gg, Buonanno:2006ui, Baibhav:2017jhs} included up to three overtones in the ringdown model, and effective-one-body (EOB) models went on to include up to seven overtones~\cite{Buonanno:2009qa}. 
Here the focus was on improving the matching of the ringdown part of the model to the merger, and the physical importance or meaning of the overtones was not discussed. 
This was, however, the focus of a 2019 work by~\citet{Giesler:2019uxc} who performed fits to NR waveforms with up to seven overtones, achieving accurate fits and recovery of remnant properties as far back as the time of peak strain.
A recent follow-up study demonstrated that up to nine overtones (alongside a number of other QNMs) can be found in the NR data with stable amplitudes, meaning a linear sum of QNMs describes the ringdown starting at times around the peak luminosity~\cite{Giesler:2024hcr}.
Prompted by this work, \citet{Forteza:2021wfq} investigated the performance of fits to NR with up to 16 overtones.
Predominantly focusing on fitting at or around the time of peak strain, it was found that sometimes as many as 12 overtones are needed for the best recovery of remnant properties.
And more recently, \citet{Oshita:2024wgt} found that fits with up to $\sim 20$ overtones were necessary to reconstruct the $(2,2)$ mode of a ringdown with a delta-function source.



Overtones, then, may have the potential to serve as a powerful tool for BH spectroscopy --- that is, the measurement of multiple QNM frequencies from which a clean and powerful test of the no-hair theorem and general relativity (GR) can be performed.
Indeed, for equal-mass, non-spinning binaries, the detection of multiple overtones may be the most promising way of performing BH spectroscopy~\cite{JimenezForteza:2020cve}, and overtones may be more sensitive to corrections to GR~\cite{Cano:2024ezp}.

Due to their potential promise for doing spectroscopy, overtones have been a natural target in ringdown analyses of the loudest events observed by the LIGO-Virgo-KAGRA network~\cite{LIGOScientific:2014pky, VIRGO:2014yos, KAGRA:2020tym}.
There was much attention on the first GW event GW150914~\cite{LIGOScientific:2016aoc} after a detection claim was made for the $(2,2,1)$ QNM~\cite{Isi:2019aib}.
This work was followed by a series of analyses reporting varied evidence for the overtone~\cite{CalderonBustillo:2020rmh, LIGOScientific:2020tif, Cotesta:2022pci, Isi:2022mhy, Finch:2022ynt, Ma:2023vvr, Crisostomi:2023tle, Isi:2023nif, Wang:2023ljx, Carullo:2023gtf, Correia:2023bfn, Wang:2024yhb, Chandra:2025ipu} (see \citet{Berti:2025hly} for a review).   
More recently, the first overtone was confidently detected in the loudest GW event to date GW250114~\cite{KAGRA:2025oiz}, with some hints of the second overtone being present (although the authors note that the second overtone is not significantly preferred at any time)~\cite{LIGOScientific:2025obp}.
No-doubt with increasing detector sensitivities there will be events for which overtones beyond the first or even second can be identified.

In this work we take a pragmatic approach and investigate the practical utility of each overtone for extracting information from the ringdown. 
Using numerical-relativity waveforms, we apply both least-squares (i.e., maximum-likelihood) and Bayesian fitting methods to assess the contribution of each overtone to the ringdown, highlighting some key features. 
We employ a simple model throughout, much in the spirit of \citet{Giesler:2019uxc}.
While we acknowledge that there are features which this model does not capture (such as time-dependent QNM amplitudes~\cite{Chavda:2024awq}, horizon/redshift modes~\cite{DeAmicis:2025xuh}, nonlinear QNMs~\cite{Cheung:2022rbm,Mitman:2022qdl}, and direct waves~\cite{Oshita:2025qmn}), it has been shown to perform surprisingly well, and is sufficient (at least as a first approximation) to start asking questions about the features of fits with large number of overtones.

We start by elucidating some characteristics of ringdown overtones in Sec.~\ref{sec:characteristics}.
In Sec.~\ref{sec:start_time} we investigate the dependence of the ringdown start time on the number of overtones included in the fit, pushing the overtone model further than in previous works.
In Sec.~\ref{sec:perturbations} we assess the ``physical content'' of each overtone by perturbing individual QNM frequencies and looking at the impact on the remnant mass and spin recovery.
Finally, in Sec.~\ref{sec:amplitudes} we employ Bayesian parameter estimation to map the correlation structures in the overtone amplitudes and phases.
A discussion and our conclusions are presented in Sec.~\ref{sec:discussion}.

\section{Overtone characteristics}
\label{sec:characteristics}

\subsection{Features of many-overtone fits}

\begin{figure*}
    \centering
    \includegraphics[width=\linewidth]{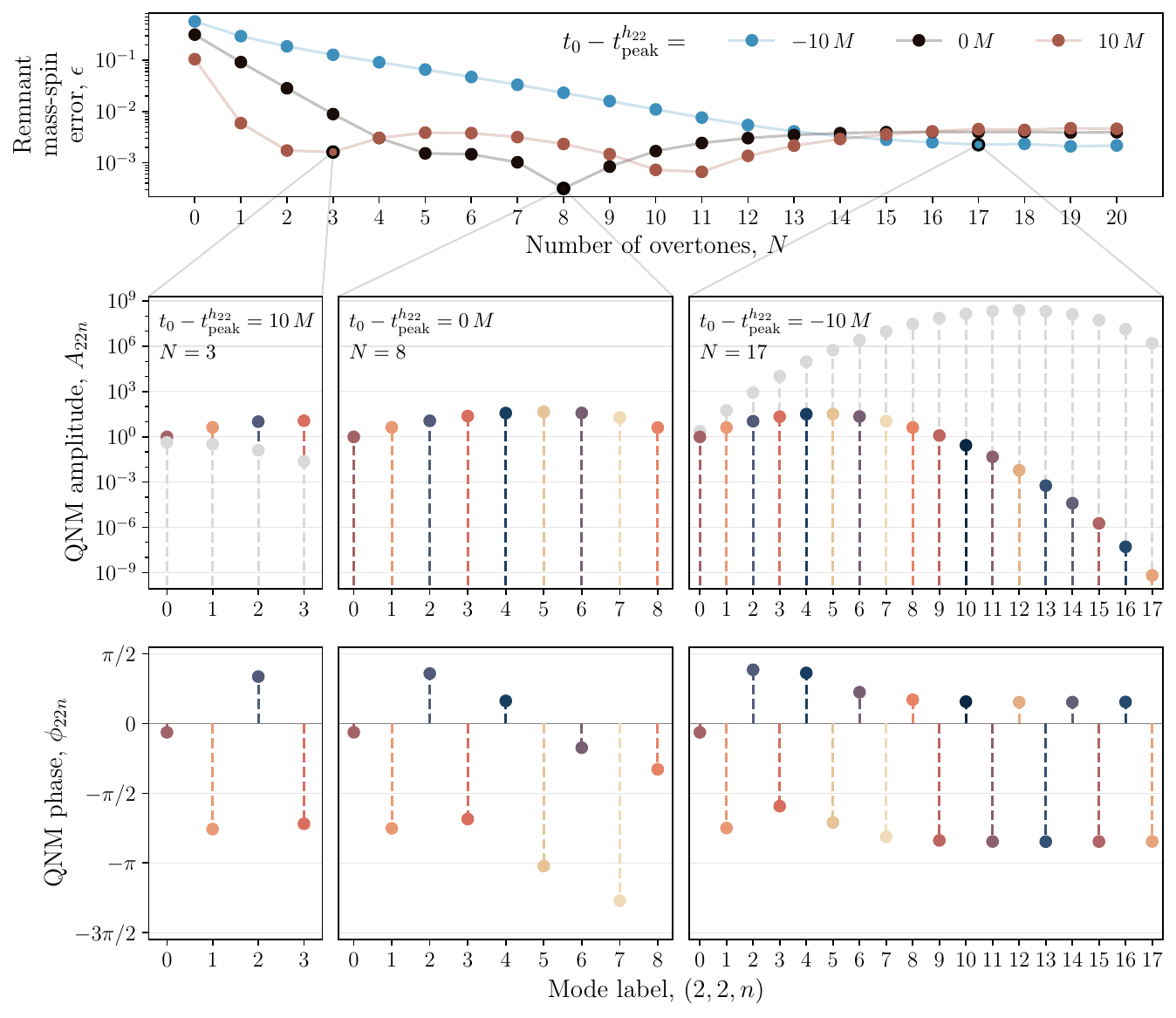}
    \caption{
    \emph{Top:} The remnant mass-spin error ($\epsilon$) from an overtone model fitted to CCE:01, for different numbers of overtones in the model ($N$) and for three different ringdown start times (line colors).
    For each start time there is a choice of $N$ which gives a minimum in the curve (indicated with a black circle --- we take the first minimum that occurs, not the global).
    \emph{Middle row:} The amplitude of each best-fit QNM from a fit at the minimum of the curve from the top panel (indicated by the connecting lines).
    We report the amplitudes rescaled to what they would be at the time $h_\mathrm{peak}^{h_{22}}$. For the left and right panels, where the fit is performed $10\,M$ after and before that time respectively, the unscaled amplitudes are shown in light gray.
    \emph{Bottom row:} The same as the middle panel, but for the phase of each best-fit QNM.
    We only show the phases reported at the time $h_\mathrm{peak}^{h_{22}}$, and the axis limits are chosen to aid comparison of the phases across the three fits.
    }
    \label{fig:even_more_overtones}
\end{figure*}

Figure \ref{fig:even_more_overtones} highlights some key features of ringdown fits with overtones.
In the figure we consider ringdown fits to the $(2,2)$ mode of an NR waveform, SXS:BBH\_ExtCCE:0001~\cite{sxs_collaboration_2024_10783245, cce_catalog} (henceforth CCE:01).
This simulation corresponds to an equal-mass non-spinning BH binary, and has been mapped to the superrest frame of the remnant BH~\cite{MaganaZertuche:2021syq, Mitman:2024uss} via the Python package \texttt{scri}~\cite{boyle_2025_15693419, Boyle:2013nka, Boyle:2014ioa, Boyle:2015nqa}.
Our ringdown model for the $(2,2)$ mode of the complex strain, $h = h_+ - ih_\times$, consists of a sum of $N$ overtones,
\begin{equation}\label{eq:h22}
    h_{22} = \sum_{n=0}^{N} C_{22n} e^{-i \omega_{22n} (t - t_0) } \qquad t \geq t_0,
\end{equation}
where $C_{\ell m n} = A_{\ell m n}e^{i\phi_{\ell m n}}$ are the complex QNM amplitudes.
In Fig.~\ref{fig:even_more_overtones} we apply this model at three different ringdown start times $t_0$: at the time of peak strain in the $(2,2)$ mode, $\tpeak$, and at $10\,M$ before and after this time.
Along with the complex QNM amplitudes, we allow the remnant mass and spin to vary in the fit and perform fits with up to $N = 20$ overtones (resulting in $2(N+1) + 2$ free parameters).
Fits are performed with the Python package \texttt{qnmfits}~\cite{qnmfits,MaganaZertuche:2025bua}, which uses linear least squares (\texttt{numpy.linalg.lstsq}~\cite{Harris:2020xlr}) to fit for the linear parameters ($C_{\ell m n}$), and a nonlinear least-squares method (in our case, \texttt{scipy.optimize.minimize}~\cite{Virtanen:2019joe}) to fit for the nonlinear parameters ($M_f$, $\chi_f$).
Further details on the fitting implementation can be found in Sec.~2.3 of Ref.~\cite{Finch:2023dcd}.
When fitting for the remnant mass and spin, we can compute an error $\epsilon$ between the best-fit remnant properties and the true remnant properties known from the simulation
\begin{equation}\label{eq:epsilon}
    \epsilon = \sqrt{\left(\frac{\Delta M_f}{M}\right)^2+(\Delta \chi_f)^2},
\end{equation}
where $\Delta M_f = M_{f,\mathrm{bestfit}} - M_{f,\mathrm{true}}$ and $\Delta \chi_f = \chi_{f,\mathrm{bestfit}} - \chi_{f,\mathrm{true}}$, and $M$ is the total binary mass.
We see from the top panel of Fig.~\ref{fig:even_more_overtones} that there is always a number of overtones $N$ for which the error $\epsilon$ reaches a minimum, after which the error starts to increase again (there may be another minimum at a larger number of overtones, but here we only consider the ``first'' minimum for reasons explained below). 
For example, at $t_0 = \tpeak$ this minimum is reached with eight overtones, comparable with the seven overtones used in \citet{Giesler:2019uxc} when studying SXS:BBH:0305~\cite{SXS:BBH:0305, Boyle:2019kee, sxs_catalog}.
When fitting at later times we see that fewer overtones are required to recover the remnant properties at comparable accuracy, consistent with the picture of higher overtones decaying away.
And when fitting $10\,M$ before the peak, we find that the recovery of the remnant properties steadily improves as we include more overtones, reaching a minimum error with 17 overtones.
As we will discuss in Secs.~\ref{sec:start_time} and \ref{sec:perturbations}, this does not mean that the ringdown starts as early as $10\,M$ before the time of peak strain; the time when a particular model appears to fit the data best depends on the exact QNM content of the model, and here we are not including a variety of subdominant QNMs (such as those which enter due to mode mixing, or quadratic QNMs) which effectively raises the noise floor of the simulation and allows us to fit earlier.

So, clearly, overtones help to accurately recover the remnant properties and allow you to start the fit to the ringdown earlier.
However, it has been argued that the overtones are simply serving to fit away anything that isn't the fundamental mode, which is the only mode actually excited.
One possible sign of this overfitting comes from the fitted overtone amplitudes and phases.
In the middle row we show the QNM amplitudes from the fits which minimize $\epsilon$ for each start time (indicated by the connecting lines); in colored markers we show the amplitudes reported at $\tpeak$, and in gray markers we show the amplitudes at the fit time $t_0$ (for the middle panel, these are the same thing).
The bottom row shows the QNM phases, for which we only report the phases at $\tpeak$. 
While the very large amplitudes achieved by the overtones are an unavoidable consequence of their rapid decay time, they could also be interpreted as a sign of overfitting.
This is particularly true when we additionally look at the phases, where a regular and apparently fine-tuned pattern emerges to achieve the required destructive interference. 
In the next subsection we take a closer look at how each overtone contributes to the ringdown and how this destructive interference is achieved.

\begin{figure*}
    \centering
    \includegraphics[width=0.9\linewidth]{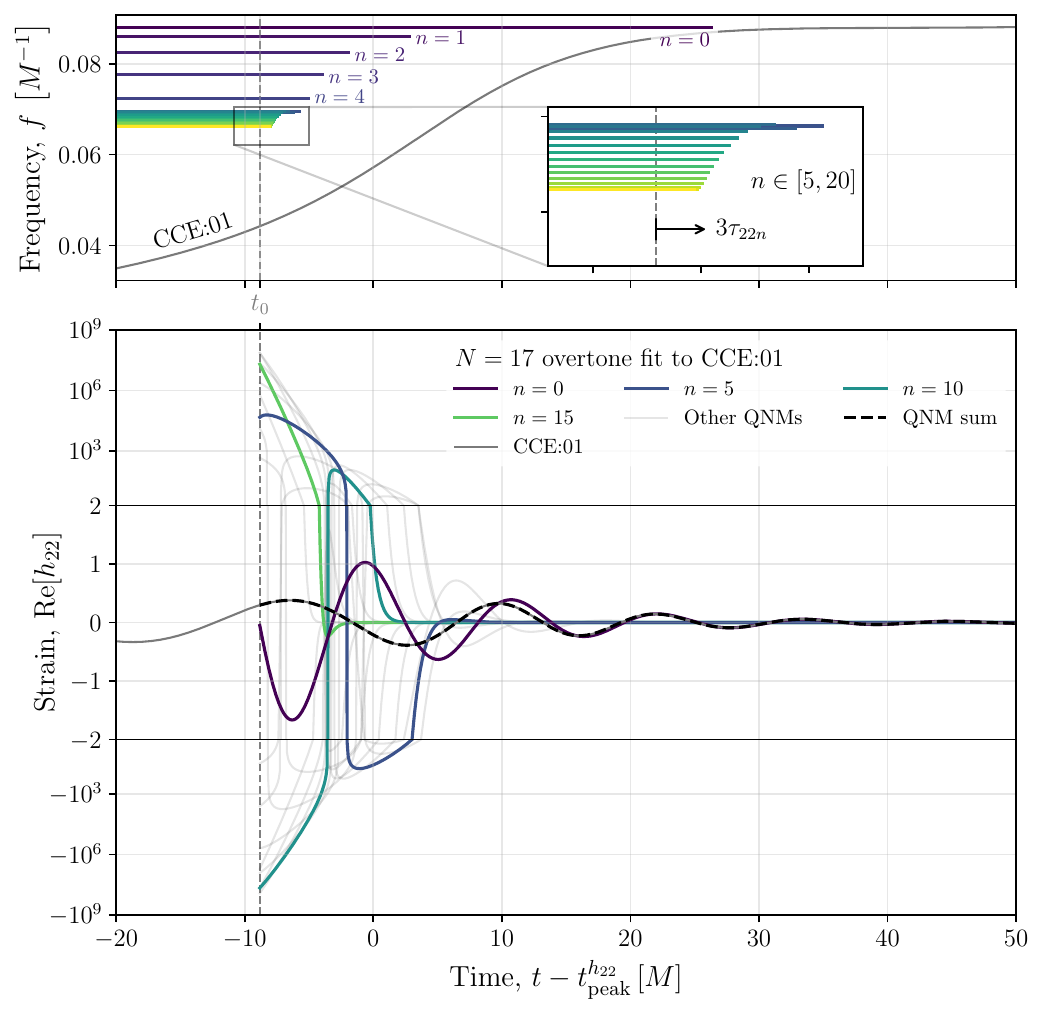}
    \caption{
    Overtone morphology.
    \emph{Top:} The frequency evolution of CCE:01 around the time of peak strain, with the overtone frequencies plotted as horizontal lines.
    The length of each line to the right of the vertical dashed line is proportional to the decay time of the mode (the line length is given by $3\tau_{22n}$). 
    \emph{Bottom:} A 17-overtone fit to CCE:01, deconstructed into the individual QNMs.
    The start time of the fit, indicated by the vertical dashed line, is determined by minimizing the mismatch (see Sec.~\ref{sec:start_time}).
    Due to the large number of QNMs in the model we highlight only a subset of them.
    As $n$ increases, the overtones grow to extremely large amplitudes (note the change in $y$-axis scale below $-2$ and above $2$) and barely oscillate before decaying away.
    }
    \label{fig:overtone_morphology}
\end{figure*}

Despite this, we do see a broad consistency in the amplitudes and phases across the different start times for these three fits; a clear sign of overfitting would be inconsistency between the measured amplitudes and phases across start times (i.e., Eq.~\ref{eq:h22} is not being followed).
The amplitudes of the fundamental and first two overtones vary by $\sim 1\%$ around their mean value across the three fits, and higher-overtone amplitudes vary by a factor $\sim 2$. 
This amplitude consistency is not guaranteed; for example, if we were to take the global minimum of the sequence of fits performed $10\,M$ after the peak (with $N=11$ overtones), we find that it prefers to give the higher overtones very large amplitudes that are completely inconsistent with the fits performed at earlier times (instead of varying by a factor $\sim 2$, we see variations up to a factor $\sim 10^9)$.
This is an example of the issues that can occur when attempting to include overtones in your model that have decayed below the noise floor (or below other unmodeled QNMs), and is a result of the fitting problem being ill-conditioned; this is the reason works such as \citet{Giesler:2024hcr} must drop overtones from their model as they fit at later times.

We emphasize that we do not see perfect amplitude and phase consistency across the three fits, but that is not surprising. 
As stated above, our model is certainly not complete, and the omission of certain features (e.g.\ subdominant QNMs, or time-dependent QNM amplitudes) will bias the result. 
And as discussed in detail in \citet{Giesler:2024hcr}, the fitting of amplitudes and phases for highly damped modes is ill-conditioned and challenging.
In Sec.~\ref{sec:amplitudes}, where we look at amplitudes in more detail, we will advocate for moving away from least-squares fits and instead introduce a likelihood with a notion of uncertainty.

\subsection{Overtone morphology}

\begin{figure*}
    \centering
    \includegraphics[width=0.9\linewidth]{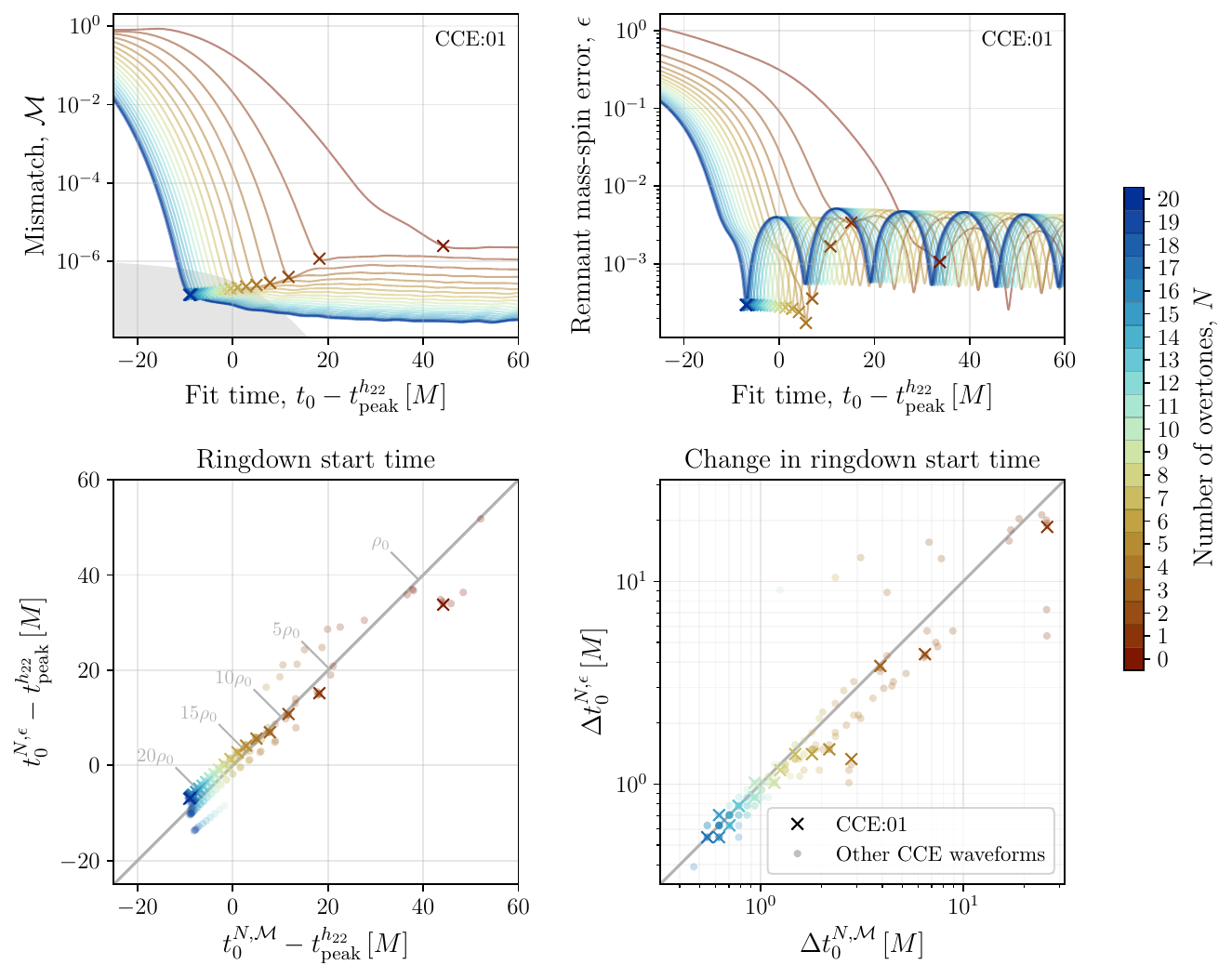}
    \caption{
    \emph{Top left:} The mismatch vs ringdown start time for fits to the $h_{22}$ mode of CCE:01 with a model consisting of $N$ overtones of the fundamental $(2,2,0)$ QNM. Crosses mark the location of $t_0^{N,\MM}$ for each $N$. The gray region indicates the NR error, obtained from the mismatch between the two highest available levels.
    \emph{Top right:} As in the top left panel, but for the error on the remnant mass and spin. Crosses mark the location of $t_0^{N,\epsilon}$ for each $N$.
    \emph{Bottom left:} The ringdown start times, $t_0^{N,\MM}$ and $t_0^{N,\epsilon}$, plotted against each other for CCE:01 (crosses) and the other nine CCE waveforms considered in this work (faint circles). We also indicate the approximate SNR (in the high-SNR limit) as a function of start time, in units of the SNR from a fundamental-mode-only fit.
    \emph{Bottom right:} As in the bottom left panel, but for the change in ringdown start time, Eq.~\ref{eq:delta_t0}. This is not defined for $N=0$, and we only plot up to $N=17$ since $\Delta t_0^{18,\MM}$ and $\Delta t_0^{18,\epsilon}$ are negative.
    }
    \label{fig:start_time}
\end{figure*}

The structure of the phases in Fig.~\ref{fig:even_more_overtones} reveals one of the key features of overtone sums in ringdown modeling; neighboring overtones tend to be out of phase with each other, and especially for large $n$ a very regular structure appears.
This is how a sum of modes with such large amplitudes results in a waveform with an amplitude $\sim \order{1}$.

To demonstrate the overtone behavior in the waveform space, in the bottom panel of Fig.~\ref{fig:overtone_morphology} we show the individual QNMs from a 17-overtone fit to CCE:01. 
Here we have fixed the remnant mass and spin to their true values known from the simulation, meaning the complex amplitudes are the free parameters (giving 36 total free parameters in this fit).
The start time of the fit was chosen to minimize the mismatch between the sum of QNMs (dashed black line) and CCE:01 (gray line) --- see Sec.~\ref{sec:start_time} for more details.
This start time roughly matches the choice of $\tpeak - 10\,M$ seen in Fig.~\ref{fig:even_more_overtones}, for which it was found 17 overtones gave the best estimate of the remnant properties.

Turning to the individual modes which constitute the QNM sum, this figure highlights the destructive interference of the overtones and their increasingly rapid decay for higher $n$.
For clarity we plot in color only a subset of the QNMs ($n = 0$, 5, 10, 15), plotting the rest with light-gray lines.
Note that in order to plot the QNM waveform for the modes with the largest amplitudes, we switch to a log scale for $\abs{\Re[h_{22}]} > 2$.
We see that the highest-$n$ mode highlighted, $n=15$, exhibits minimal oscillatory behavior (at least on the linear scale) as it rapidly decays, and on our scale becomes consistent with zero at around the time of peak strain.
Lower-$n$ modes ``switch-off'' at successively later times (and complete more visible cycles), until about $20\,M$ after the peak only the fundamental mode remains.
Due to the alternating phases we see the large amplitudes contributing both positively and negatively to the overall sum, which is shown with the dashed-black line.
Although overtones are most strongly correlated with neighboring-$n$ modes, we see that they are to some extent correlated with all other overtones --- this is the focus of Sec.~\ref{sec:amplitudes}.
Indeed, if we isolate, say, the $n=11$ and $n=12$ overtones from the fit (which have amplitudes of $\approx 9.15 \times 10^7$ and $1.05 \times 10^8$, respectively, at $\tpeak - 10\,M$), despite being out of phase, their sum still has an amplitude $\approx 1.39 \times 10^7$.
It is only when all the modes in the fit are summed together that amplitudes of $\sim \order{1}$ are achieved.

The top panel of Fig.~\ref{fig:overtone_morphology} shows the frequency evolution of CCE:01 over time (gray line), with the frequencies of the $\ell = m = 2$ QNMs (with $n$ from 0 to 20) indicated with horizontal lines.
We also indicate the decay time of the mode via its length: each line has a length $3 \tau_{22n}$ from $t_0$.
At late times we see the CCE:01 frequency tend to the fundamental-mode frequency, consistent with the bottom panel.
Overtones then enter at lower frequencies, tracking the ``chirp'' of the merger-ringdown (and no-doubt allowing the overtones to fit to the early-time behavior).
The frequency dependence on $n$ is not entirely monotonic: for CCE:01 ($\chi_f \approx 0.69$) the $n=7$ mode has a slightly higher frequency than $n=6$ (see the inset --- the modes can be identified by their line lengths, which by definition order them).
The frequencies then decrease monotonically again until $n=19$.
Using data available from Ref.~\cite{cook_2025_16801267}, we checked the overtone behavior up to $n=33$ and see that the real part of the frequency starts to oscillate around $f \sim 0.0633\,M^{-1}$.
The details of the spectrum will depend on the BH spin --- Fig.~\ref{fig:taxonomy} plots the complex overtone frequencies as a function of spin. 


\section{Ringdown start time}
\label{sec:start_time}

One useful aspect of overtones is that they appear to push the linear ringdown description (Eq.~\ref{eq:h22}) to earlier times in the signal.
From a data-analysis perspective, this allows us to start our ringdown analysis earlier, and maximize the signal-to-noise ratio (SNR).
However, clearly each additional overtone can't result in an SNR gain as high as the one before, so in some sense there are ``diminishing returns'' from each successive overtone.
In this section we explore this by identifying a suitable ringdown start time, and seeing how this varies, for each choice of overtone number $N$.
The ringdown start times we quote in this section are not necessarily ``true'' start times --- we are simply finding an optimal time (either in terms of the waveform match or the recovery of remnant properties) and identifying this as a suitable choice of ringdown start time from a practical perspective.

We first employ the mismatch to identify the ringdown start time, which is essentially a measure of the overlap between two waveforms.
We define the mismatch, $\MM$, between two complex timeseries $h_1$ and $h_2$ as 
\begin{equation}
    \MM = 1 - \frac{\langle h_1 | h_2\rangle}{\sqrt{\langle h_1 | h_1\rangle\langle h_2 | h_2\rangle}},
\end{equation} 
where the inner product $\langle h_1 | h_2\rangle$ is taken to be 
\begin{equation}
    \langle h_1 | h_2\rangle = \Re\left[\int_{t_0}^{t_0 + T} \dd{t} h_1(t) \, h_2^*(t) \right]
\end{equation} 
(a star denotes complex conjugation).
A low mismatch means there is a good overlap between the two waveforms, and so by computing the mismatch between a NR waveform and a best-fit ringdown model as a function of the ringdown start time $t_0$ we can investigate at what time the model appears to fit the data best.

Starting with CCE:01 (again working with only the $\ell = m = 2$ spherical harmonic), we compute the mismatch between a best-fit ringdown model and CCE:01 for a range of ringdown start times $t_0$ and number of overtones $N$.
The QNM frequencies are fixed to their true values via the remnant mass and spin in these fits.
We also fix $T$, the duration of the mismatch integral, to be $100\,M$.
We show the resulting mismatch curves in the top left panel of Fig.~\ref{fig:start_time}.
For a given $N$, the mismatch curves have a distinctive shape where they rapidly fall at early times before flattening out.
The initial rapid decrease in mismatch is due to content in the data which is not included in our model, but is rapidly decaying as we move to later times (whether that be higher overtones, nonlinearities near merger, or some other component). 
So, this slope is indicative that we are fitting too early (i.e., before a suitable ringdown start time for that choice of $N$).
The flattening out of the curve is due to the presence of the $(3,2,0)$ mode in the data, which we are not modeling here. 
This mode (and, in principle, all modes with $m = 2$) is present in $h_{22}$ due to mode mixing~\cite{Berti:2014fga}.
In effect, the $(3,2,0)$ is acting like a noise floor in our data, and where the mismatch curve flattens out can be identified as where the highest overtone in the model (with $n = N$) falls below the $(3,2,0)$ mode.
Of course, the mismatch is an integral over time, so really the ``knee'' of the mismatch curve corresponds to something more like where the power of the highest overtone falls below the power in the $(3,2,0)$ mode.
Regardless, the mismatch curves offer us a clean way of identifying a suitable start time as a function of $N$ (if we were to instead include the $(3,2,0)$ mode in our fit, it becomes harder to identify an appropriate start time --- see Appendix~\ref{app:with_320}). 
To identify a suitable ringdown start time we employ a knee-finding algorithm~\cite{5961514} as implemented in the \texttt{kneed} Python package~\cite{arvai_2023_8127224}, which essentially finds the point of maximum curvature.
We indicate these knees with crosses in the top left panel of Fig.~\ref{fig:start_time}, where we see the trend where each additional overtone moves the knee to earlier times.
It turns out that this only works up to a point; in fact, for all CCE waveforms considered in this work, we see the mismatch knee does not occur at an earlier time when moving from $N=17$ to 18 (which is why we chose $N=17$ in Fig.~\ref{fig:overtone_morphology}).
This may be due to the accuracy of the NR data or the fitting method (we would expect that with more free parameters, a better fit should be obtained).
For the highest numbers of overtones considered here the shift in start time is becoming comparable to the simulation time resolution (with time steps of $\sim 0.08$), which may also be a factor.
We also note that the fit does improve again when moving to $N=19$ overtones, becomes worse for 20, and improves again for 21; it may be that the neighboring overtones become so strongly correlated that they must be added in pairs for the fits to continue to improve.

The fact that including more degrees of freedom in our model results in a better fit is not particularly surprising, so as a check we can additionally allow the remnant mass and spin (which determine the QNM frequencies) to vary in the fit and see how well we recover the true values.
A suitable choice of start time, as well as fitting the data well, should recover the remnant mass and spin.
As before, we can quantify this via the error $\epsilon$, given in Eq.~\ref{eq:epsilon}.
A large $\epsilon$ implies we are applying our model in a regime where it is not valid, since the QNM frequencies must be modified from their true values in order to get a good fit to the data.
In the top-right panel of Fig.~\ref{fig:start_time} we show how $\epsilon$ varies as a function of ringdown start time for a range of different overtone numbers, for CCE:01.
As expected, at sufficiently early times there is a large bias in the remnant mass and spin recovery, but as for the mismatch curves we see that the fits perform better at later times, and that each additional overtone achieves a comparable accuracy earlier.
Instead of flattening out we see a periodic pattern at late fit times, which we have verified is due to the presence of an unmodeled $(3,2,0)$ mode in the data (this structure changes when we include the $(3,2,0)$ mode in the fit --- see Appendix~\ref{app:with_320}).

As for the mismatch, we can use the $\epsilon$ curves to identify a suitable ringdown start time, which we simply take to be the time of the first minimum in the curve.
These are indicated with the crosses.
These minima appear at similar times to the mismatch-curve knees, and we plot them against each other in the bottom left panel of Fig.~\ref{fig:start_time}.
On the $x$-axis we plot the start times as obtained from the mismatch curves, $t_0^{N,\mathcal{M}}$, and on the $y$-axis we plot the start times as obtained from the remnant mass-spin error curves, $t_0^{N,\epsilon}$.
In this panel we also show the results from all the aligned-spin simulations in the publicly available CCE catalog (specifically we exclude the simulations with IDs 8, 9, and 13 --- see Ref.~\cite{cce_catalog} for the properties of the simulations).
Although it is hard to quantify how close to the diagonal line is ``good enough'' (where the diagonal line is perfect agreement between $t_0^{N,\MM}$ and $t_0^{N,\epsilon}$), one way to gauge the bias is by calculating $\epsilon$ at the lowest-mismatch time $t_0^{N,\MM}$ instead of the lowest-error time $t_0^{N,\epsilon}$. 
When we do this, we see maximum values of $\epsilon$ of $\sim 4 \times 10^{-3}$ across all simulations and all values of $N$ (not substantially higher than the lowest values obtained for CCE:01).
The agreement between the two methods is reassuring (and consistent with Fig.~\ref{fig:even_more_overtones}), and gives us confidence that our chosen ringdown start times are optimal for both the quality of the fit and the recovery of the remnant properties.

On the bottom-left panel we also indicate the approximate SNR gain as a function of start time for CCE:01, to give an idea of the contribution of each additional overtone.
We perform a simple calculation, working only with $h_+$, and assuming that the ringdown occurs over a sufficiently narrow range of frequencies for the PSD to be approximated as a constant.
We work in units of $\rho_0$,
\begin{equation}
    \rho_0^2 \approx \alpha \int_{\bar{t}_0^{N=0}}^\infty |h_+(t)|^2 ~ \mathrm{d}t,
\end{equation}
where $\bar{t}_0^{N=0}$ is the average of $t_0^{N=0,\MM}$ and $t_0^{N=0,\epsilon}$, and $\alpha$ is a constant that contains the PSD, antenna response, and any mass or distance scaling applied to the CCE waveform.
We then adjust the lower-limit of the integration to find the ringdown start time with a particular SNR (say, $5 \rho_0$).
The key point here is that the SNR gain from zero to two overtones is as big as the SNR gain in going from two to 20 overtones --- this what was meant by the idea of diminishing returns mentioned above.

Finally, on the bottom-right panel of Fig.~\ref{fig:start_time} we show the change in start time as determined by the two methods,
\begin{equation}\label{eq:delta_t0}
    \Delta t_0^N = t_0^{N} - t_0^{N-1},
\end{equation}
where we only show results up to $N=17$, since $\Delta t_0^{18}$ is negative.
Across all simulations considered we see the general trend that higher overtones result in smaller changes to the start time.
This is just another way of expressing the diminishing returns of the overtones, except expressed directly in terms of the ringdown start time.
This is suggestive of some asymptotic limit to how early a sum of overtones can model the ringdown --- perhaps with higher-resolution data and robust fitting methods, an arbitrary number of overtones can be extracted, each with less of a contribution than the one before (in other words, there is no clear maximum or highest overtone excited in the fit, and it is just a matter of data quality and fitting method which number can be extracted).
What the ``true'' asymptotic limit of the start time is remains an open question; we emphasize that the start times quoted here should not be interpreted as absolute ringdown start times, since we are neglecting QNM content and effectively raising our noise floor (which results in the optimal start time occurring at earlier times --- see Appendix~\ref{app:with_320}).
A more careful fitting procedure, such as in \citet{Giesler:2024hcr}, finds that times only as early as the peak luminosity (generally a few $M$ after the time of peak strain) can be reached --- this time may be more indicative of when a linear sum of QNMs becomes a valid description.

\section{Frequency perturbations}
\label{sec:perturbations}

Ultimately, a key motivation for studying the ringdown is that it provides a clean and powerful test of GR. 
The program of BH spectroscopy relies on measurements of multiple QNM frequencies, and as mentioned above this is already a reality for ringdown overtones~\cite{Isi:2019aib,KAGRA:2025oiz}.
However, as for the ringdown start time, there is some expectation that, due to their increasingly rapid decay, higher overtones may be less ``useful'' for performing spectroscopy. 
In other words, any deviations from Kerr in the overtone spectrum may become increasingly harder to detect for each subsequent overtone, meaning even if we had sufficient SNR to detect many overtones this may not actually provide a meaningful test of the Kerr metric and the no-hair theorem.
In this section we investigate this by introducing perturbations to the overtone complex frequencies, and then perform fits to the ringdown where we allow the remnant mass and spin to vary. 
Different to previous works, we will be perturbing overtone frequencies one at a time to gauge the individual contributions of each overtone.
Any bias in the recovered mass and spin, quantified via $\epsilon$ (Eq.~\ref{eq:epsilon}), is indicative of the ``physical content'' of the overtones and their capacity for performing meaningful BH spectroscopy. 

\begin{figure}
    \centering
    \includegraphics[width=\linewidth]{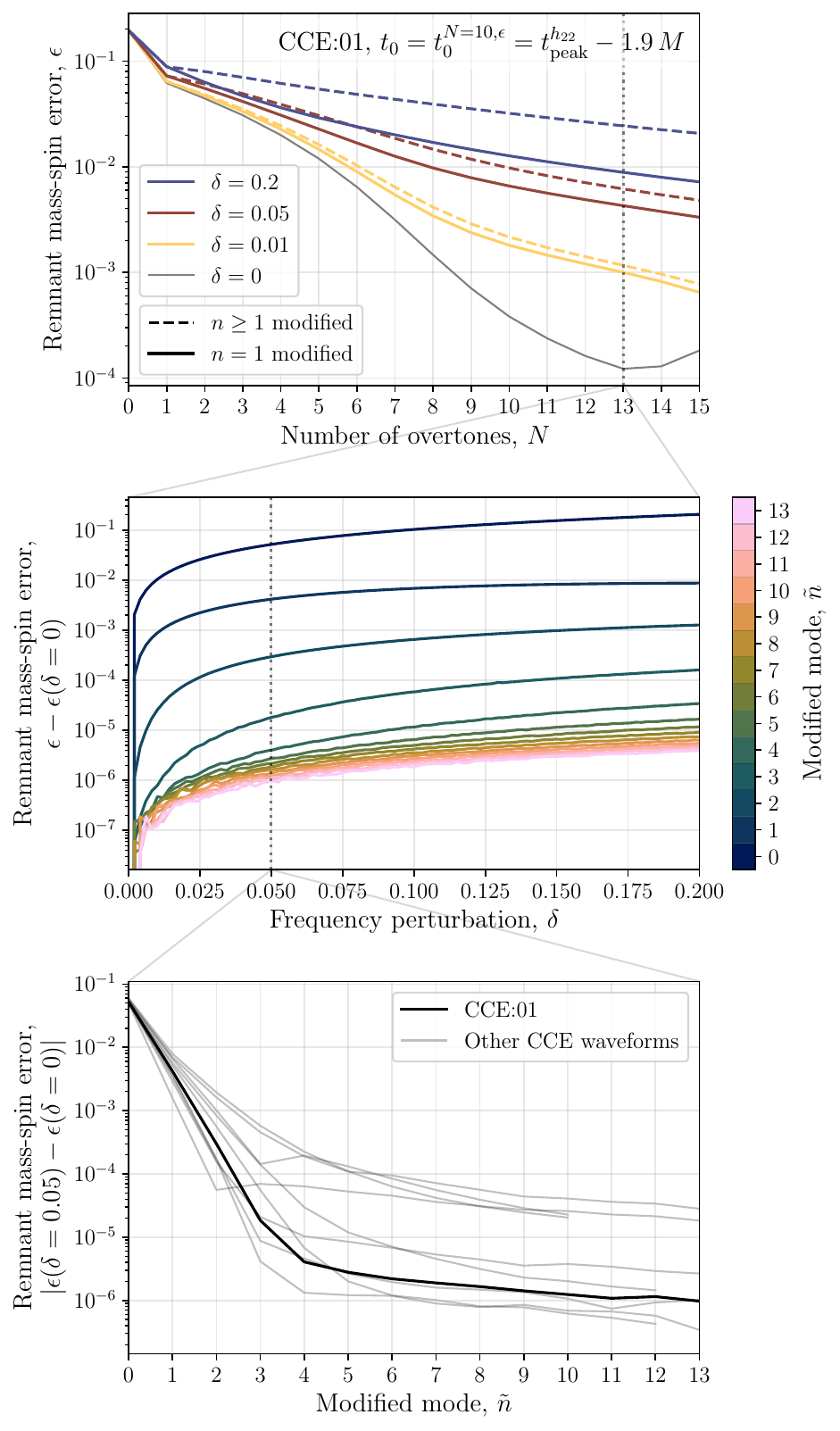}
    \caption{
    \emph{Top:} The induced error on the remnant mass and spin as a function of the number of overtones included in the fit, when perturbing all the overtone frequencies simultaneously (dashed lines) and when perturbing only the $n=1$ mode (solid lines), for different values of the perturbation $\delta$ (line colors). 
    We fit to CCE:01 at a start time $t_0 = t_0^{N=10,\epsilon}$ (as determined in Sec.~\ref{sec:start_time}), but now we include the $(3,2,0)$ mode in the fit.
    \emph{Middle:} The remnant error for a $N=13$ overtone fit as a function of $\delta$, where now we modify each overtone frequency individually in turn (line colors).
    \emph{Bottom:} A slice through the middle panel at $\delta = 0.05$.
    Alongside CCE:01 we plot the result from each of the CCE waveforms considered in this work.
    }
    \label{fig:frequency_perturbations}
\end{figure}

Previous works have performed similar studies.
In particular, Fig.~7 of \citet{Giesler:2019uxc} was used to demonstrate that the overtones are indeed physically meaningful. 
We reproduce this figure in the top panel of Fig.~\ref{fig:frequency_perturbations}.
For demonstration purposes, we take CCE:01 and perform fits at a start time of $t_0^{N=10,\epsilon} \approx \tpeak - 1.9\,M$ (this is the time which minimizes $\epsilon$ for a 10-overtone fit; see Sec.~\ref{sec:start_time}).
In this section we choose to include the $(3,2,0)$ mode in our model, as we find it improves the interpretability of the results but we do not ever perturb its frequency. 
As in \citet{Giesler:2019uxc}, we plot the error on the remnant mass and spin, $\epsilon$, as a function of the number of overtones included in the fit, $N$.  
The gray line shows the result for the unperturbed QNM frequencies. 
Notice that this line does not have a minimum at $N=10$, despite us performing the fit at $t_0^{N=10,\epsilon}$; this is because we are including the $(3,2,0)$ QNM in our fit, whereas in Sec.~\ref{sec:start_time} we neglected this mode.
As a result, we effectively lower our ``noise floor'', and consequently we can include more overtones before overfitting.
This clearly demonstrates that the start times obtained in Sec.~\ref{sec:start_time} are contingent on the exact model used.
We then introduce perturbations to the frequencies, which we parameterize through $\delta$; the modified QNM frequency is given by
\begin{equation}\label{eq:delta}
    \tilde{\omega}_{\ell m n} = (1 + \delta)\, \omega_{\ell m n},
\end{equation}
where $\omega_{\ell m n}$ is the GR value.
\citet{Giesler:2019uxc} chose to perturb all overtones simultaneously, leaving only the fundamental mode with the GR value.
We show this with the dashed lines in the top panel of Fig.~\ref{fig:frequency_perturbations}.
Clearly a bias is incurred when we perturb the frequencies, and this grows with $\delta$ in agreement with \citet{Giesler:2019uxc} (except now pushed to a higher number of overtones).

One can conclude that there is something special about the overtone frequencies, i.e., it is not the case that any highly damped modes can be used to fit away everything that isn't the fundamental mode.
Since we are modifying all overtones together, this is a statement about the overtones collectively.
But, we are also interested in the importance of overtones individually.
To this end, in the top panel of Fig.~\ref{fig:frequency_perturbations} we also include the result when only the $n=1$ overtone has its frequency modified (solid line).
The resulting values of $\epsilon$ are comparable to those obtained when modifying all overtones, clearly indicating that the first overtone is largely responsible for the incurred bias.
The bias from modifying only the first overtone is less than when modifying all the overtones, so the higher overtones do still have some contribution to the inference of the remnant mass and spin (just less than the fundamental mode).

To explore this further, in the middle panel of Fig.~\ref{fig:frequency_perturbations} we modify the frequency of each overtone in turn, and plot the resulting $\epsilon$ relative to $\epsilon(\delta = 0)$ as a function of the perturbation size $\delta$.
We again fit to CCE:01, and we use the same ringdown start time as in the top panel. 
As indicated by the connecting lines we use the $N = 13$ model (a natural choice, since this model results in the lowest remnant error at this start time when we include the $(3,2,0)$ mode).
The main takeaway here is the diminishing impact of the overtones on the remnant property error as we go to higher overtone number.
To highlight this, in the bottom panel of Fig.~\ref{fig:frequency_perturbations} we take a slice at $\delta = 0.05$.
For the first four overtones we see that the bias is roughly an order of magnitude less than the bias from perturbing the overtone before, and then for higher $\tilde{n}$ we see a pile-up at $\epsilon(\delta = 0.05) - \epsilon(\delta = 0) \sim 10^{-6}$.
We note that the intrinsic simulation error on the remnant mass and spin, obtained by comparing $M_f$ and $\chi_f$ for the two highest available levels for CCE:01, is $\sim 10^{-5}$.
However, the behavior in the middle panel is also seen when working with a mock NR waveform built from a sum of overtones with exact Kerr frequencies, so we believe this is a genuine feature caused by the overtone morphology.
In the bottom panel we also show the result for the other aligned-spin CCE waveforms, which show the same trend.
For each CCE waveform we use the appropriate $t_0^{N=10,\epsilon}$, and then perform the fits with the value of $N$ which minimizes $\epsilon$ in the unperturbed case (which is not necessarily 13 as it was for CCE:01; this is why the lines on the bottom panel are of different lengths).
Note that we take the absolute value for clarity, since for CCE:05 and CCE:11 $\epsilon$ falls below the unperturbed $\epsilon(\delta = 0)$ value when certain overtone frequencies are modified.

Other works which have performed similar analyses, such as \citet{Baibhav:2023clw}, have come to the conclusion that above some value of $n$ the overtones are unphysical (in the case of Ref.~\citet{Baibhav:2023clw}, they claim that the $n \geq 2$ are unphysical).
We instead come to the conclusion that there is no clear ``highest overtone``, and the number of overtones included in the model is just a matter of SNR and the precision required. 
Of course, practically speaking the SNRs required to detect a bias in the remnant mass and spin at the level of $\sim 10^{-6}$ will not be achieved with current generation detectors, and it may even be out of reach of next generation detectors (a rough Fisher-matrix calculation suggests SNRs at least on the order of 1000 are required).
Nonetheless, we see no reason to impose a cut at some value of $n$ for what is regarded physical.

We have verified that the same picture holds at start times earlier and later than $t_0^{N=10,\epsilon}$ (just with more and less overtones required, respectively). 
However, for earlier times and more overtones, we run into increasingly noisy estimates of $\epsilon$.
We suspect this is due to a combination of our minimization routine and strong correlations between overtones; we explore the latter further in Sec.~\ref{sec:amplitudes}.

\section{Amplitudes and phases}
\label{sec:amplitudes}



An important piece of information not readily available from the least-squares fits are the correlations between neighboring overtone amplitudes and phases.
To investigate this further we will introduce a likelihood and a notion of uncertainty into the fits, thereby allowing us (with a choice of prior) to obtain posterior distributions and to map the correlations.
This is in a similar spirit to \citet{Clarke:2024lwi} (and, more recently, \citet{Dyer:2025iwj, Dyer:2025hdt}), but, as we will explain, we have a different implementation and choice for the noise (see also \citet{Redondo-Yuste:2023seq} who perform Bayesian fits to NR).

\subsection{Likelihood}

Our data is the $(2,2)$ mode from a NR simulation, which consists of two timeseries $d_+$ and $d_\times$ sampled at $K$ times $d_{+/\times} = [d_{+/\times}(t_0),\,d_{+/\times}(t_1),\, \ldots,\, d_{+/\times}(t_{K-1})]$. 
We will refer to NR data with the symbol $d$, and our ringdown model with $h$, and we will suppress the $22$ label in this section for clarity.
We adopt a white-noise Gaussian likelihood with constant standard deviation $\sigma$, and we treat $d_+$ and $d_\times$ as independent ``measurements'' to obtain a overall likelihood
\begin{multline}\label{eq:likelihood}
    \log{\mathcal{L}(d|h,\theta)} = \log{\mathcal{L}_+(d_+|h_+,\theta)} + \log{\mathcal{L}_\times(d_\times|h_\times,\theta)} \\
    = -\frac{1}{2} \sum_{k=0}^{K-1} \left[ \frac{d_k^+ - h_k^+(\theta)}{\sigma} \right]^2 + (+ \leftrightarrow \times) + \mathrm{const}
\end{multline}
where $\theta$ is our vector of parameters (in our case, the QNM amplitudes and phases).
Of course, there isn't actually a statistical uncertainty associated with our data; rather, there is a deterministic truncation error that arises from numerically solving the Einstein field equations~\cite{Kopriva_2009,Szilagyi:2014fna,Boyle:2019kee} (with some additional error contributions, for example from the procedure used to extract the GWs from the simulation~\cite{Boyle:2009vi,Zlochower:2012fk,Chu:2015kft}).
Unfortunately it is non-trivial to estimate this error, with the most common method being to compare the output of simulations ran at two different resolutions (this was the approach adopted by Refs.~\cite{Dyer:2025iwj, Dyer:2025hdt}).
It is also worth noting that when we perform a least-squares fit (a well established method in the literature), we are making exactly the same assumption as Eq.~\ref{eq:likelihood} about the ``measurement uncertainty'' of our data.
The only difference is that a least-squares fit returns the parameters which maximize Eq.~\ref{eq:likelihood} (that is, they are the maximum-likelihood values).

We must now also pick a value for our noise level, $\sigma$. 
In this work we take a conservative approach; we choose to intentionally leave the $(3,2,0)$ mode out of our model (as in Eq.~\ref{eq:h22}), and treat it as an effective noise floor. 
Then, we can use the $(3,2,0)$ amplitude to set $\sigma$.
The key idea is that the $(3,2,0)$ is the ``loudest'' component of the data we can reliably fit for which is not a $(2,2,n)$ QNM (motivated by the results shown in Fig.~\ref{fig:start_time}, and Appendix~\ref{app:with_320}).
We could, alternatively, choose to model the $(3,2,0)$ mode, but then our effective noise floor will presumably be the next dominant QNM.
However, reliably extracting the amplitude of subdominant QNMs becomes challenging, and we found that the $(3,2,0)$ QNM was the easiest to work with for the simulations considered in this work.
Another approach would be to attempt to model all QNMs, which would mean the noise floor is simply the error of the simulation.
However, this error is hard to quantify, and since we are primarily interested in the behavior of the overtones we choose to avoid modeling other subdominant QNMs.

\begin{figure}
    \centering
    \includegraphics[width=\linewidth]{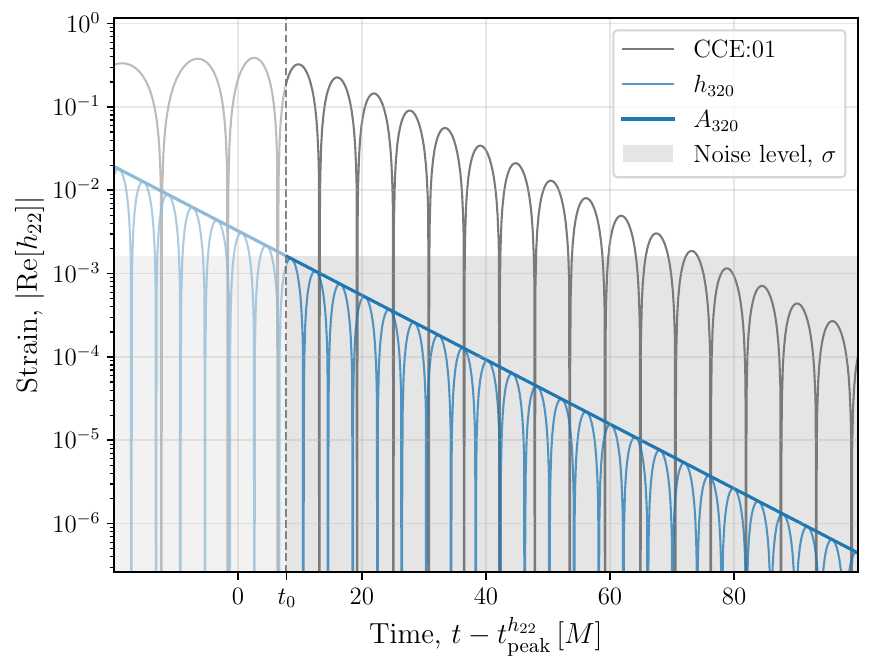}
    \caption{
    How we pick the noise level in this work.
    Given the $(2,2)$ mode of an NR waveform (here we show CCE:01, thin gray line), we perform a fit of the $(3,2,0)$ QNM over a stable window to reconstruct $h_{320} = A_{320}e^{i\phi_{320}}e^{-i\omega_{320}t}$ thin blue line).
    We then choose the noise level, $\sigma$ (gray shaded region) to be equal to $A_{320}$ (thick blue line) at the chosen $t_0$. 
    }
    \label{fig:noise}
\end{figure}

Our exact procedure is as follows. 
We first perform a least-squares fit to the $h_{22}$ data at a range of ringdown start times, using a model consisting of only the $(2,2,0)$ and $(3,2,0)$ QNMs.
At sufficiently late times any overtones have decayed, and the amplitudes and phases of the $(2,2,0)$ and $(3,2,0)$ modes stabilize.
From the series of fits performed at different start times, we identify a window of size $10\,M$ over which the variance in the amplitude and phase of the $(3,2,0)$ QNM is minimized --- we use the mean value of the amplitude and phase over this window as our fitted value (this is a similar procedure as performed in other works, for example see~\citet{Cheung:2023vki}).
From the fitted values of $A_{320}$ and $\phi_{320}$ we can reconstruct the $h_{320}$ contribution to $h_{22}$ as a function of time (given by $h_{320} = A_{320}e^{i\phi_{320}}e^{-i\omega_{320}t}$), which we show in Fig.~\ref{fig:noise}.
In the figure the gray line shows the $(2,2)$ mode of CCE:01 (absolute value of the real part), and the thin blue line shows our reconstructed $h_{320}$ mode. 
The thick blue line shows just the amplitude $A_{320}$.
Although $A_{320}$ (and $\phi_{320}$) were extracted from a particular stable window of times (we found that generally the modes stabilized $60\, M$ after the time of peak strain), we can use the known frequency and decay time to extrapolate the mode to any time.
Of course, as discussed above, we don't expect this extrapolation to work to arbitrarily early times, so we consider this a first approximation to the evolution of the mode.
Finally, we simply pick our noise level $\sigma$ to be equal to $A_{320}$ at whatever ringdown start time $t_0$ we are using (meaning the exact value of $\sigma$ used depends on $t_0$).
In the figure, $t_0$ (which is just an example) lies between 0 and $20\,M$ after the time of peak strain, and the gray band indicates the value of $\sigma$ used.
Via analysis of mock signals (whereby the QNM amplitudes and phases were known exactly) we found this choice for $\sigma$ was conservative enough to recover the injected amplitudes at $90\%$ credibility.
We experimented with making $\sigma$ time dependent such that it followed $A_{320}$ (i.e., uncorrelated heteroscedastic Gaussian noise), however we encountered issues due to the late time behavior when the $(3,2,0)$ mode decays away (effectively up-weighting the late-time data).
To avoid this one could impose a secondary noise floor at late times which becomes dominant when the $(3,2,0)$ mode has decayed, but we want to avoid picking arbitrary noise floors and opt to have $\sigma$ constant in time.

To obtain the posterior, we make use of the fact that the complex QNM amplitudes enter the model linearly. 
Then, since we have a Gaussian likelihood, if we also have Gaussian priors on the complex QNM amplitudes then our posterior is Gaussian and can be written in a closed form (see, for example, \citet{Hogg:2020jwh} and Refs.~\cite{Dyer:2025iwj, Dyer:2025hdt}, and note that this approach is also used in other codes such as Ref.~\cite{maximiliano_isi_2024_13892015, Isi:2021iql}).
Following the notation of Ref.~\cite{Hogg:2020jwh} we can write our model as
\begin{equation}
    h_{+/\times} = M_{+/\times} \cdot \theta,
\end{equation}
where $\theta$ is our parameter vector, length $2(N+1)$:
\begin{equation}
    \theta = \left(\mathrm{Re}[C_0],\ \mathrm{Im}[C_0],\ \ldots,\ \mathrm{Re}\left[C_N\right],\ \mathrm{Im}\left[C_N\right]\right)^\mathsf{T}
\end{equation}
and $M_{+/\times}$ are our design matrices (using ``c'' and ``s'' as shorthand for cos and sin, respectively):
\begin{widetext}
\begin{equation}
    M_+ = \begin{bmatrix} 
         e^{-t_0/\tau_0}\mathrm{c}(\omega_0 t_0) & e^{-t_0/\tau_0}\mathrm{s}(\omega_0 t_0) & e^{-t_0/\tau_1}\mathrm{c}(\omega_1 t_0) & e^{-t_0/\tau_1}\mathrm{s}(\omega_1 t_0) & \cdots & e^{-t_0/\tau_N}\mathrm{c}(\omega_N t_0) & e^{-t_0/\tau_N}\mathrm{s}(\omega_N t_0) \\ 
         e^{-t_1/\tau_0}\mathrm{c}(\omega_0 t_1) & e^{-t_1/\tau_0}\mathrm{s}(\omega_0 t_1) & e^{-t_1/\tau_1}\mathrm{c}(\omega_1 t_1) & e^{-t_1/\tau_1}\mathrm{s}(\omega_1 t_1) & \cdots & e^{-t_1/\tau_{N}}\mathrm{c}(\omega_{N} t_1) & e^{-t_1/\tau_{N}}\mathrm{s}(\omega_{N} t_1) \\
         \vdots & \vdots & \vdots & \vdots & \ddots & \vdots & \vdots \\
         e^{-t_{\tilde{K}}/\tau_0}\mathrm{c}(\omega_0 t_{\tilde{K}}) & e^{-t_{\tilde{K}}/\tau_0}\mathrm{s}(\omega_0 t_{\tilde{K}}) & e^{-t_{\tilde{K}}/\tau_1}\mathrm{c}(\omega_1 t_{\tilde{K}}) & e^{-t_{\tilde{K}}/\tau_1}\mathrm{s}(\omega_1 t_{\tilde{K}}) & \cdots & e^{-t_{\tilde{K}}/\tau_{N}}\mathrm{c}(\omega_{N} t_{\tilde{K}}) & e^{-t_{\tilde{K}}/\tau_{N}}\mathrm{s}(\omega_{N} t_{\tilde{K}})
         \end{bmatrix} \nonumber
\end{equation}
\begin{equation}
    M_\times = M_+ (\cos \rightarrow \sin,\ \sin \rightarrow -\cos)
\end{equation}
\end{widetext}
which have shape $ K\ \mathrm{(timesteps)} \times 2(N+1)\ \mathrm{(parameters)} $. 
We defined $\tilde{K} = K - 1$ for brevity.
We can write our overall likelihood as
\begin{equation}
    \mathcal{L} = \prod_{j \in \{+,\times\}} \mathcal{N}\left( d_j | M_j \cdot \theta, C_j \right)
\end{equation}
where, as discussed above, the noise covariance $C_j = \sigma^2 I_K$ ($I_K$ is a $K \times K$ identity matrix).
Bayes' theorem can then be written as (\citet{Hogg:2020jwh} Eq.~28)
\begin{equation}
    \mathcal{N}(\theta|\mu,\Lambda) ~ \mathcal{L} = \mathcal{N}(\theta|a,A) ~ \prod_{j\in\{+,\times\}}\mathcal{N}(d_j|b_j,B_j).
\end{equation}
The Gaussian $\mathcal{N}(\theta|\mu,\Lambda)$ is our prior on $\theta$, and $\mathcal{N}(\theta|a,A)$ is our posterior on $\theta$. 
In this work we take the limit $\Lambda \rightarrow \infty$ (resulting in an uninformative, uniform prior on the real and imaginary parts of the complex amplitudes). 
We can then solve for $a$ and $A$ with the following iteration:
\begin{gather}
    A_0^{-1} = \Lambda^{-1} = 0 \qquad  x_0 = \Lambda^{-1} \cdot \mu = 0 \\
    A_j^{-1} = A_{j-1}^{-1} + M_j^\mathsf{T} \cdot C_j^{-1} \cdot M_j \\
    x_j = x_{j-1} + M_j^\mathsf{T} \cdot C_j^{-1} \cdot h_j \\
    a_j = A_j \cdot x_j,
\end{gather}
finishing with $A = A_J$ and $a = a_J$.
There are also expressions for the $b_j$ and $B_j$, which give us the Bayesian evidence, but these are not used here.

\subsection{Measurement correlations}

\begin{figure*}
    \centering
    \includegraphics[width=\linewidth]{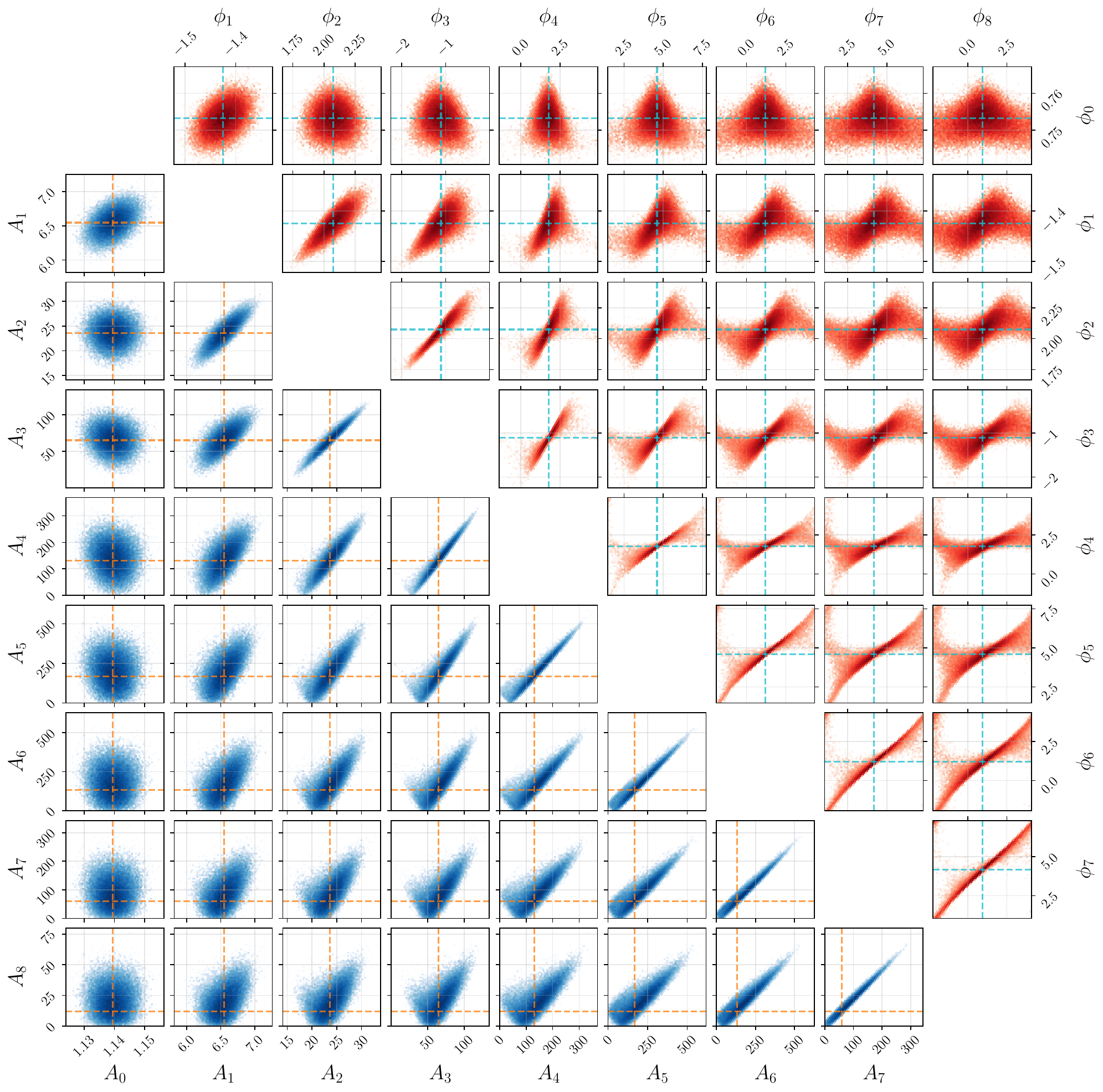}
    \caption{
    Correlation structures in the amplitudes (lower triangle) and phases (upper triangle) from a eight-overtone fit to CCE:01, with a ringdown start time $t_0 = t_0^{N=8,\mathcal{M}} \approx \tpeak - 1.7\,M$.
    Dashed lines indicate the result from a least-squares fit with the same model and start time.
    }
    \label{fig:correlations}
\end{figure*}

Given the above framework we can draw samples from a known Gaussian posterior without any concerns related to convergence, meaning we can reliably explore correlation structures in the amplitudes and phase posteriors.
This builds on previous works that have looked at overtone correlations (see, for example, \citet{Bhagwat:2019dtm}) --- here we look at the full structure in the posterior and extend to higher overtones.
We show an illustrative example in Fig.~\ref{fig:correlations}, which consists of an eight-overtone fit to CCE:01 with a ringdown start time $t_0 = t_0^{N=8,\mathcal{M}} \approx \tpeak - 1.7\,M$.
At this start time, the chosen value of the noise level was $\sigma \approx 4 \times 10^{-3}$.
In the figure we show the posterior density on the amplitudes (lower triangle) and phases (upper triangle), with darker regions corresponding to regions of higher posterior density.
Dashed lines indicate the result from a least-squares fit with the same ringdown model and start time.
It is important to note that our prior is flat on the real and imaginary parts of $C_n = A_n e^{i\phi_n}$, but here we plot the posterior in $A_n$ and $\phi_n$. 
The effective prior in this space is proportional to $A_n$ for the amplitudes and is flat for the phases, such that higher amplitudes are favored.

Along the main diagonal of Fig.~\ref{fig:correlations} we see increasingly strong correlations between neighboring overtone amplitudes and phases as we go to higher $n$.
This structure is a result of the destructive interference exhibited by the overtones (See Figs.~\ref{fig:even_more_overtones} and \ref{fig:overtone_morphology}); the amplitudes of each overtone must increase together in order for their sum to remain the same (likewise, their phases must change together to retain the destructive interference).
Through a ``domino'' effect we see that overtones are also correlated with modes beyond their immediate neighbors, however we see that the strength of the correlations diminish for overtones further apart in $n$.
It is also interesting to note that the direction of the degeneracy is different for each pair of overtones (particularly for the amplitudes --- note the axis scales).
This degeneracy (i.e., the parameter we are actually measuring) is presumably some combination of amplitudes weighted by the mode frequency and decay time.

\begin{figure*}
    \centering
    \includegraphics[width=0.9\linewidth]{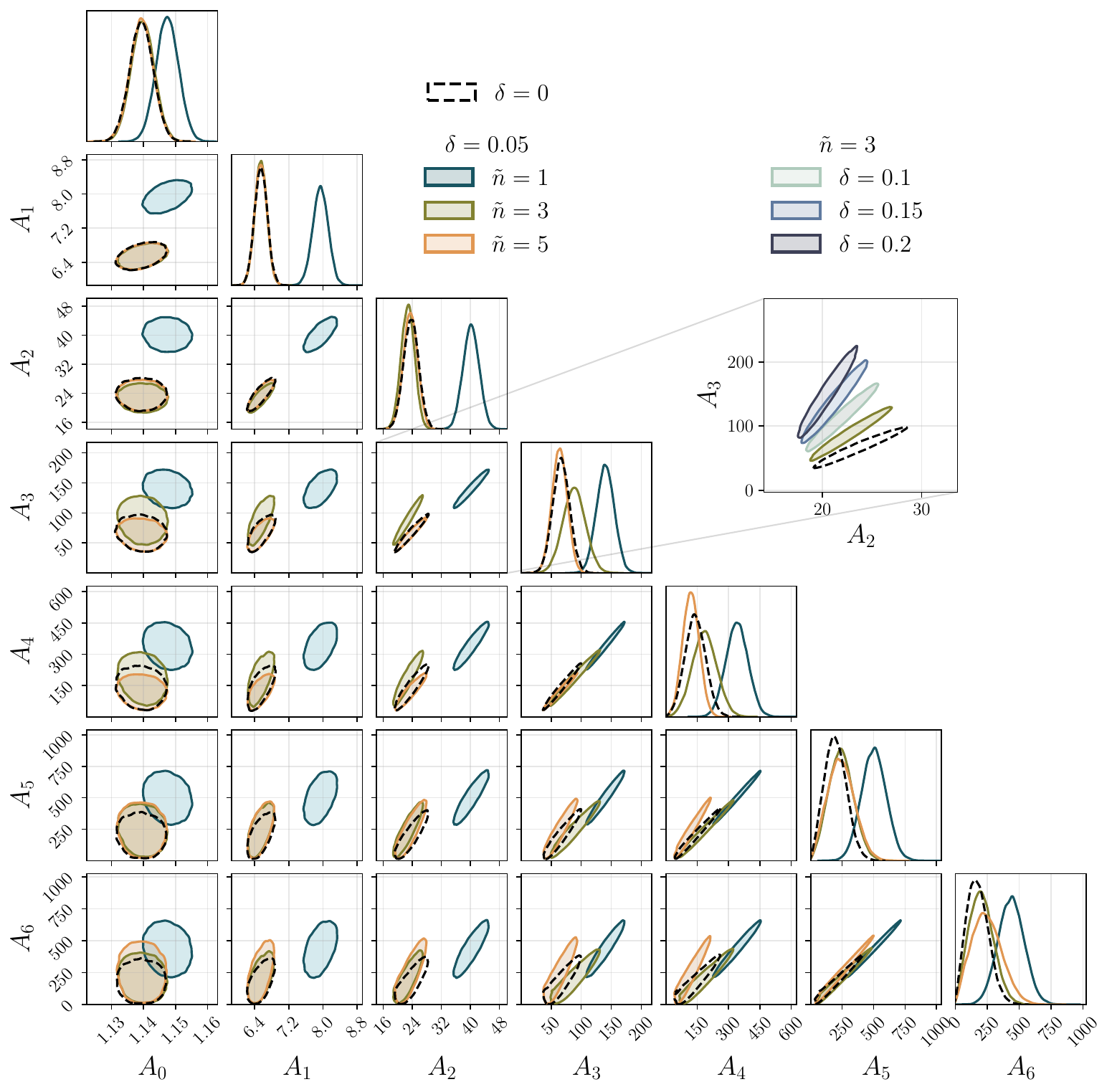}
    \caption{
    The impact of perturbing select overtone frequencies on the amplitude posteriors, for the same fit as in Fig.~\ref{fig:correlations} (we only show posteriors up to the sixth overtone for clarity).
    In the main corner plot we use a fixed perturbation size, $\delta = 0.05$, and modify the first (blue), third (green), and fifth (orange) overtone frequencies.
    In the inset we consider the effect of different perturbation sizes, focusing on perturbation to the third overtone and plotting the joint posterior between $A_3$ and $A_2$.
    The dashed black contours correspond to the unperturbed posterior, which is the same as in Fig.~\ref{fig:correlations}.
    All contours correspond to 90\% credible regions.
    }
    \label{fig:bayesian_perturbations}
\end{figure*}

We have verified that the strength of this degeneracy continues to grow for higher numbers of overtones, and is consistent across all the CCE simulations considered in this work.
Such correlations could pose problems for parameter estimation and sampling the posterior; here we have used flat priors on the real and imaginary parts of the complex amplitudes to obtain a closed-form expression for the posterior, but that is not possible if we wanted to use flat priors on the amplitudes $A_n$. 
Re-weighting is one option, however we find the number of effective samples drop with each additional overtone, and beyond $\sim$ five overtones reweighting becomes infeasible. 

Beyond parameter estimation, these correlations could cause problems for building models of QNM amplitudes.
In the absence of a first-principles understanding of QNM excitation, we rely on fits to NR waveforms to extract the QNM amplitudes and phases needed to construct predictive models~\cite{Cheung:2023vki, MaganaZertuche:2024ajz, Mitman:2025hgy}.
However, for higher overtones it becomes increasingly challenging to extract the amplitudes and phases with certainty, which could hinder this approach --- another way in which the utility of each overtone decreases with $n$.

Finally, the conservative noise level used here and the choice of prior (which up-weights larger amplitudes) will act to broaden and stretch the posteriors, emphasizing the degeneracies. 
While clearly the degeneracies are intrinsic to the overtones (it is some combination of overtone amplitudes and phases that we measure best), further study will be needed to understand the impact of the noise level and prior.

\subsection{Impact of frequency perturbations}

In Sec.~\ref{sec:perturbations} we considered the effect of frequency perturbations on the recovery of the remnant mass and spin, finding that each successive overtone has less impact.
However, this ignored any effect of the perturbations on the amplitudes and phases.
As discussed above, measurements of individual overtone amplitudes and phases are made challenging due to strong correlations --- so, in this section we look at the impact of frequency perturbations on the correlation structure instead.

Our main findings are summarized in Fig.~\ref{fig:bayesian_perturbations}.
We perform the same fit as in Fig.~\ref{fig:correlations}, but now introduce a perturbation to individual overtone frequencies (see Eq.~\ref{eq:delta}).
Note that we perform the fits with the mass and spin fixed to their true values; as shown in Sec.~\ref{sec:perturbations} a different mass and spin may fit the data better when we introduce perturbations to the frequencies, and we have checked that using the ``best-fit'' (as determined by a least-squares fit) mass and spin does not impact our conclusions.
In the main corner plot of Fig.~\ref{fig:bayesian_perturbations} plot 90\% credible regions on the QNM amplitudes when we perturb only the first ($\tilde{n} = 1$, blue), third ($\tilde{n} = 3$, green), and fifth ($\tilde{n} = 5$, orange) overtones with a fixed perturbation size $\delta = 0.05$.
For reference, the dashed black lines show the unperturbed posterior (this is the same posterior as in Fig.~\ref{fig:correlations}).
Although the fit included eight overtones, for clarity we only show six.
We find that perturbations to the longer lasting (i.e., low-$n$) mode frequencies tend to ``destabilize'' the amplitudes of all modes in the fit.
This can be seen for the $\tilde{n} = 1$ posteriors in the figure, where all amplitudes are significantly shifted compared to the $\delta = 0$ case. 
There are larger and smaller shifts for the $\tilde{n} = 0$ and $\tilde{n} = 2$ cases respectively (not shown).
Perturbations to the third overtone frequency result in amplitude shifts for all modes with higher $n$ (which makes intuitive sense, since these modes ``overlap'' the most), but the largest impact is to neighboring overtone amplitudes.
This picture continues to hold for higher $\tilde{n}$, with the impact on amplitudes becoming increasingly localized around neighboring overtones.
We also see that modifications to higher $n$ modes seem to have little impact on the one-dimensional marginal distributions --- the impact of the frequency perturbations are most noticeable in the joint posteriors.
For example, for the $\tilde{n} = 5$ case we see a clear feature in the joint $A_4$-$A_5$ posterior, despite the one-dimensional posteriors largely overlapping with the $\delta = 0$ case for these modes.

We highlight this for the $\tilde{n} = 3$ case with the inset plot, which shows the joint $A_2$-$A_3$ posterior for different choices of $\delta$. 
The dashed-black and green contours are the same as in the main corner (corresponding to $\delta = 0$ and 0.05 respectively), and in addition we add posteriors for $\delta = 0.1$, 0.15, and 0.2.
The structure of the correlation changes depending on the frequency shift; so, although individual overtone amplitude posteriors provide less constraining GR consistency tests as we go to higher $n$ (due to the fact that they become individually less well measured), it may be that the joint amplitude measurements provide a more sensitive test (even as we go to high $n$).

Finally, we note that beyond the first few overtones, frequency perturbations seem to have little impact on the phase posteriors.
It seems that a particular arrangement of phases is strongly favored for $n$ greater than three or four, and they resist changes (at least for the systems and perturbations considered here).

\section{Discussion and Conclusions}
\label{sec:discussion}

Overtones have emerged as a promising target in the ringdown of binary BH mergers, potentially allowing analyses to start earlier in time (increasing the available ringdown SNR) and allowing tests of the no-hair theorem via measurements of multiple ringdown QNMs. 
However, fits to the ringdown with overtones have some arguably peculiar features, primarily driven by their rapid decay time, and it remains an open question whether overtones are excited to the extent that fits to NR would suggest.
After elucidating some of these features (Figs.~\ref{fig:even_more_overtones} and \ref{fig:overtone_morphology}), in this work we take a pragmatic approach and ask, assuming the ringdown can be modeled as a sum of many overtones, what each additional overtone is actually contributing in terms of the ringdown start time (Fig.~\ref{fig:start_time}) and ability to test the no-hair theorem (Fig.~\ref{fig:frequency_perturbations}).

Contrary to previous claims in the literature, we argue that there is no clear maximum or highest relevant overtone, and instead the number of overtones included in the model is simply a matter of signal SNR.
That said, one of the main messages of this work is that each successive overtone (despite increasingly large amplitudes) is actually less useful from a practical perspective for extracting SNR and for doing tests of the no-hair theorem.
It is also worth mentioning that including additional modes will tend to increase model complexity, and so even if including extra overtones increases the SNR it is possible the Occam penalty will still dominate and result in minimal gains in the Bayesian evidence~\cite{CalderonBustillo:2020rmh, Chandra:2025ipu}.
So, whilst increasingly more overtones may be included in ringdown models as we detect higher SNR signals, one should be aware of what each overtone is contributing to the model and what constraints each overtone is actually providing.
To investigate the ringdown start time and ability to test the no-hair theorem we have worked within the simplified framework of least-squares fits, which has allowed us to see general trends of fits with large numbers of overtones and make some approximate statements about the signal SNR required for such fits to be useful.
A more complete picture could be obtained via full Bayesian parameter estimation with high-overtone models and high-SNR mock ringdown signals, where all parameters (including deviation parameters, $\delta f$ and $\delta \tau$) are sampled over.

\begin{figure*}
    \centering
    \includegraphics[width=0.9\linewidth]{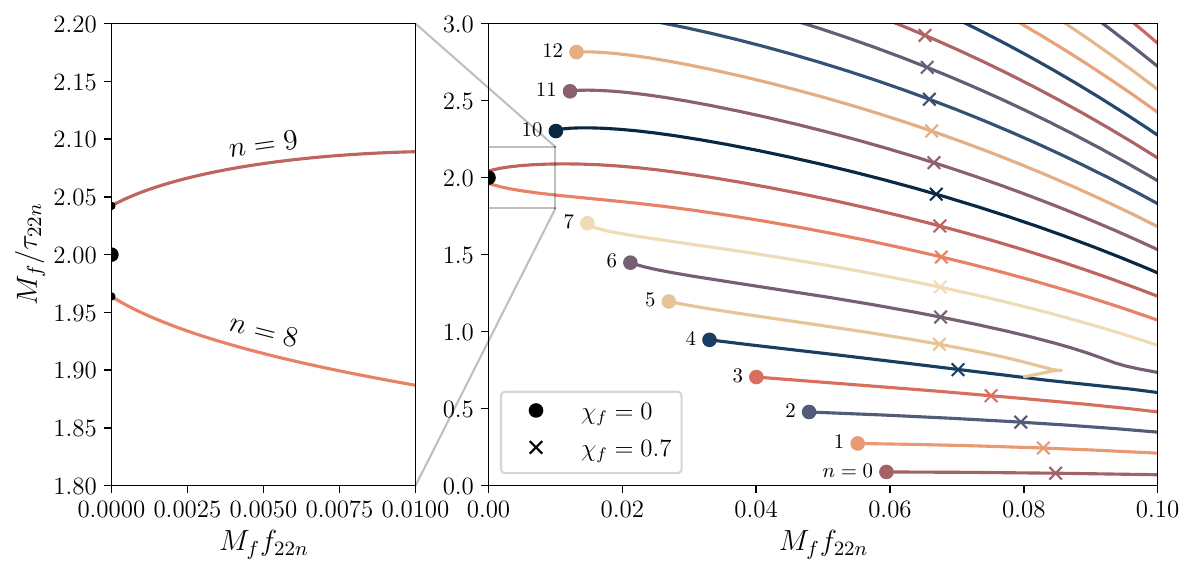}
    \caption{
    The labeling convention used for the overtones in this work.
    We show the frequency and decay time (scaled by the remnant BH mass) for the $\ell = m = 2$ family of modes, for the first few overtones.
    Circle markers indicate the Schwarzschild ($\chi_f = 0$) spectrum (for which the $m$ index has no meaning).
    Note that the $n = 8$ overtone has zero real part.
    For Kerr BHs, the points become lines which we parameterize by the spin.
    For example, we show the spectrum of a $\chi_f = 0.7$ BH with the crosses.
    Two $m = 2$ frequencies appear to emerge from the Schwarzschild $\ell = 2$, $n = 8$ frequency.
    We choose to label these the $(2,2,8)$ and $(2,2,9)$ QNMs, but note that different choices have been made in the literature.
    }
    \label{fig:taxonomy}
\end{figure*}

As a first step in this direction we move beyond the commonly used least-squares fits to NR, and employ Bayesian parameter estimation to investigate the correlations that exist between the overtones (Fig.~\ref{fig:correlations}).
Another method of testing GR with the ringdown is by performing amplitude and phase consistency tests, and here we find overtones also present some challenges.
Due to strong correlations between neighboring overtones, the measurement of individual overtone amplitudes and phases becomes increasingly challenging for higher $n$.
However, we find that the joint measurement of overtone amplitudes is more sensitive to their frequencies (Fig.~\ref{fig:bayesian_perturbations}), perhaps offering an avenue for performing consistency tests (i.e., a consistency test with some better-measured combination of overtone amplitudes).
Of course, this would rely upon the correlation structure of the overtone amplitudes being predictable from the progenitor BH properties, and a study obtaining posteriors on the overtone amplitudes for a range of systems would be another direction for future work.
And, again, since here the remnant mass and spin were fixed for the analysis, in future sampling should be done over the full parameter space to obtain the full correlation structure.

The code used to perform all the analyses and create all figures is available at Ref.~\cite{code}.


\begin{acknowledgments}

We thank Max Isi, Will Farr, Christopher Moore, Richard Dyer, William Yang, Katerina Chatziioannou and Saul Teukolsky for helpful comments and discussions.
This work was supported by NSF Grant
PHY-2150027 as part of the LIGO Caltech REU Program which funded E.C.
E.F. acknowledges support from the Department of Energy under award number DE-SC0023101.
The authors are grateful for computational resources provided by the LIGO Laboratory and supported by National Science Foundation Grants PHY-0757058 and PHY-0823459.
This material is based upon work supported by NSF's LIGO Laboratory which is a major facility fully funded by the National Science Foundation.

\end{acknowledgments}


\appendix

\section{Overtone labeling conventions}\label{app:taxonomy}

In the Schwarzschild limit, the $\ell = 2$, $n=8$ overtone is an algebraically special mode which lies on the imaginary axis~\cite{MaassenvandenBrink:2000iwh}.
First seen by \citet{Berti:2003jh}, for non-zero spin a pair of QNMs are associated with this mode for both $m=1$ and $m=2$.
We show this in Fig.~\ref{fig:taxonomy} for the $m=2$ case, where circle markers indicate the Schwarzschild spectrum of modes, and lines show the paths traced by the $\ell = m = 2$ frequencies as spin is increased.
The eighth Schwarzschild overtone appears to discontinuously split into two branches.
For this reason, these two branches have been referred to as overtone ``multiplets''~\cite{Cook:2014cta, Cook:2016ngj}, labeled as $(2,2,8_0)$ and $(2,2,8_1)$.
However, in this work we deal with Kerr BHs (with spins $\sim 0.7$), and from a practical perspective these two branches behave as $n=8$ and $n=9$ overtones. 
So, as indicated by Fig.~\ref{fig:taxonomy}, we label them as such (this also follows the convention of \citet{Forteza:2021wfq}).

\begin{figure*}
    \centering
    \includegraphics[width=0.9\linewidth]{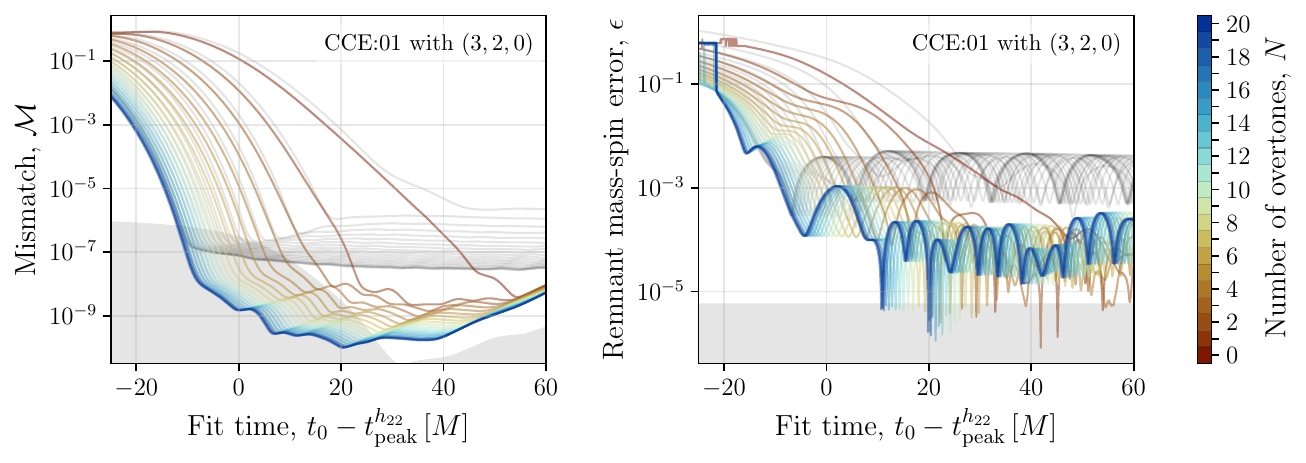}
    \caption{
    Mismatch and error on the remnant mass and spin vs ringdown start time for fits to the $h_{22}$ mode of CCE:01 with a model consisting of $N$ overtones of the fundamental $(2,2,0)$ QNM, plus the $(3,2,0)$ QNM. 
    The gray region indicates the NR error, obtained from the mismatch between the two highest available levels. 
    For reference, gray lines show the result from the top panel of Fig.~\ref{fig:start_time}, where the $(3,2,0)$ QNM was not included.
    }
    \label{fig:mm_eps_with_320}
\end{figure*}

\section{Including the $(3,2,0)$ mode}\label{app:with_320}

In Secs.~\ref{sec:start_time} and \ref{sec:amplitudes} we chose to omit the $(3,2,0)$ QNM from our model.
In Sec.~\ref{sec:amplitudes} the motivation was to use the $(3,2,0)$ mode as an effective noise floor (we can easily fit for the $(3,2,0)$ mode amplitude and use that to set a conservative noise level --- see Fig.~\ref{fig:noise}).
In Sec.~\ref{sec:start_time} it was because omitting the $(3,2,0)$ mode helps us to pick an optimal ringdown start time cleanly.
We demonstrate this in Fig.~\ref{fig:mm_eps_with_320}, where we reproduce the top row of Fig.~\ref{fig:start_time}, but now including the $(3,2,0)$ QNM in all of our fits.
We obtain lower mismatches and lower remnant mass and spin errors; by including the $(3,2,0)$ QNM we have effectively lowered our noise floor, and (as discussed in Sec.~\ref{sec:perturbations}) the optimal start times will be shifted later in time.
The new noise floor will presumably be the next loudest QNM, some other feature in the waveform, or perhaps contributions from the NR error.
But, it becomes less clear what an optimal start time is when we include the $(3,2,0)$ QNM.
Any flattening of the mismatch curves becomes less well defined, and the remnant mass-spin error exhibits more complex behavior.

\bibliography{bibliography}

\end{document}